# Metabolizable Bi$_2$Se$_3$ Nanoplates: Biodistribution, Toxicity, and Uses for Cancer Radiation Therapy and Imaging

*Xiao-Dong Zhang, Jie Chen, Yuho Min, Gyeong Bae Park, Xiu Shen, Sha-Sha Song, Yuan-Ming Sun, Hao Wang, Wei Long, Jianping Xie, Kai Gao, Lianfeng Zhang, Saijun Fan, Feiyue Fan, and Unyong Jeong\**

[*] Dr. X. D. Zhang, Dr. J. Chen, X. Shen, S. S. Song, Y. M. Song, S. Fan, F. Fan A. B.
Tianjin Key Laboratory of Molecular Nuclear Medicine, Institute of Radiation Medicine, Chinese Academy of Medical Sciences and Peking Union Medical College, Tianjin, 300192, China,

[*] Prof. U. Jeong, Y. Min, K. Park.
Department of Materials Science and Engineering, Yonsei University, 134 Shinchon-dong, Seoul, Korea,
Email: E-mail:ujeong@yonsei.ac.kr

Prof. L. Zhang, K. Gao
Key Laboratory of Human Disease Comparative Medicine, Ministry of Health, Institute of Laboratory Animal Science, Chinese Academy of Medical Sciences & Comparative Medical Center, Peking Union Medical College, Beijing 100021, China

Prof. J. Xie
Department of Chemical and Biomolecular Engineering,
National University of Singapore, 10 Kent Ridge Crescent, 119260, Singapore

**Abstract**. Bi, a high atom number element, has a high photoelectric absorption coefficient, and Se element has anticancer activity. Hence, their compound chalcogenide ($Bi_2Se_3$) deserves a thorough investigation for biomedical applications. This study reveals that $Bi_2Se_3$ nanoplates (54 nm wide) protected with poly(vinylpyrollidone) (PVP) are biocompatible and have low toxicity even at a high dose of 20 mg/kg in mice. This conclusion was made through the studies on the biodistribution and 90-day long term *in vivo* clearance of the nanoplates. Liver and spleen were dominant organs for the nanoplates accumulation which was mainly due to RES absorption, but 93 % the nanoplates were cleared after 90 days treatment. Concentrations of Bi and Se in tumor tissue continuously increased until 72 h after intraperitoneal injection into mice. Such selective accumulation of Bi was utilized to enhance the contrast of X-ray CT images. The Bi elements concentrated in a tumor led to damage on the tumor cells when exposed to gamma radiation. Growth of the tumor significantly delayed and stopped after 16 days after the tumor was treated with the $Bi_2Se_3$ nanoplates and radiation. This work clearly shows that the $Bi_2Se_3$ nanoplates may be used for cancer radiation therapy and CT imaging. They deserve further studies for biological and medical applications.

## 1. Introduction

Bismuth selenide ($Bi_2Se_3$) has attracted wide interest in condensed matter physics due to the high thermoelectric figure of merit and the unique surface electronic states as a topological insulator.[1-7] Its biomedical applications, however, have rarely been explored. Se is a natural anticancer element which can inhibit reactive oxygen species.[8] Bi is a high atom number material (Z=83) and it has a high photoelectric absorption coefficient (5.3 $cm^2/g$ at 100 keV) which is higher than those of I, Au, and Pt. The large X-rays absorption is advantageous for the use as cancer radio sensitizer and X-ray contrast agent.[9-13] Thus, $Bi_2Se_3$ nanoparticles deserve a thorough biomedical investigation for the high radiation enhancement effect and the cancer radiation therapy.

Interaction between nanostructured materials and biological system is closely related to surface chemistry.[14-23] Nanoparticles without a protected organic layer are agglomerated immediately in blood due to the strong interaction with serum proteins.[13,24] The size of the particle also strongly influences the mechanism of clearance in body.[23,25-30] It is well documented that nanoparticles during long time circulation in blood can be continuously accumulated in tumor cells through the enhanced permeability and retention (EPR) of the cancerous cells.[30-35] Coating with poly(vinylpyrrolidone) (PVP) on nanoparticles have been reported to prolong the circulation, and such nanoparticles have been successfully utilized in X-ray imaging,[36] MRI imaging,[37] and photothermal cancer therapy.[38]

Investigation of the toxicity and clearance of new materials is important for possible medical applications.[39-41] Clearance routes through liver and kidney are crucial in metabolism of materials. Nanoparticles smaller than 5.5 nm can be cleared rapidly by renal clearance,[42] while nanoparticles in the range of 10-50 nm can be absorbed by the first-pass extraction from the reticuloendothelial system (RES).[43,44] Nanoparticles larger than 50 nm can be cleared partially by the liver metabolism.[45-48] For example, a recent toxicological investigation in monkeys showed that 52 nm CdSe quantum dots encapsulated with phospholipids micelle could be partially cleared after 90 days treatment.[49] Thus, searching metabolizable nanomaterials with efficient biological activity and evaluating their biological response are important for medical application of the materials.

In this work, the PVP-coated $Bi_2Se_3$ nanoplates (NPs, 53.8 nm in width and less than 6 nm in thickness) were employed to investigate their integration in tumor cells and the possibility of the use in the cancer radiation therapy and the X-ray computed tomography (CT) imaging. The $Bi_2Se_3$ NPs intraperitoneally injected in mice exhibited continuous accumulation in the tumor cells. The Bi elements in tumor significantly enhanced the tumor radiation. Moreover, 90 days long-term toxicological response showed that 93 % $Bi_2Se_3$ NPs can be metabolized, and no any toxicity were found. This work suggests a new way of cancer therapy and CT imaging with $Bi_2Se_3$ NPs.

## 2. Results and discussions
### 2.1 Analysis of the PVP-stabilized Bi2Se3 nanoplates

Previously, we have reported the formation of ultra-thin single crystal $Bi_2Se_3$ nanodiscs and nanosheets in the presence of PVP.[50] We modified the synthetic path to decrease the size of the NPs. Stoichiometric ratio of sodium selenite ($Na_2SeO_3$) and bismuth nitrate pentahydrate ($Bi(NO_3)_3·5H_2O$) were dissolved in ethylene glycol with mild stirring. At 180 °C, reduction was triggered by rapid injection of an ethylene glycol solution of hydrazine hydrate. Details are found in the experimental section. The as-synthesized NPs were two dimensional plates (Figure 1a,b). $Bi_2Se_3$ preferentially forms a two dimensional layered structure. Consecutive five atomic layers stack along the *c*-axis ($Se_1$–Bi–$Se_2$–Bi–$Se_1$) which is a quintuple layer (QL~1 nm).[51] PVP can bind to basal plane composed of negatively charged elemental Se. Average lateral size of the NPs was ~ 53.8 nm from TEM images (Figure S1a) which is consistent with the hydrodynamic diameter (59.8 nm) obtained by dynamic light scattering (Figure S1b). High-resolution TEM images show the crystal structure of the NPs (Figure 1c,d). The hexagonal lattice fringes with a lattice spacing of 0.207 nm (Figure 1c) represent the lattice spacing of the (110) plane. The cross-section image (Figure 1d) shows a lattice spacing of 0.478 nm which corresponds to the (006) plane. Powder X-ray diffraction (XRD) confirms that the $Bi_2Se_3$ NPs have the rhombohedral crystal structure with lattice constants of $a$=0.4139 nm and $c$=2.8636 nm (Figure 1e). The NPs in this study show a good agreement with the peaks of bulk $Bi_2Se_3$ with the same crystal structure indicated by the blue vertical lines.[52] The NPs were stabilized by PVP surfactant so that they were dispersed well in water. The zeta potential of the NPs in pure water was -16.69 mV, which is in a regime of stable dispersion. We tested the time-dependent stability of the NPs, and found that hydrodynamic diameter was stable and no significant aggregation was occurred in the phosphate-buffered saline (PBS) after 1 month (Figure S1b). Inspired by these results, we tested the biological response of the $Bi_2Se_3$ NPs in mice.

## 2.2 Biodistribution and clearance of the $Bi_2Se_3$ nanoplates

Knowledge on the biodistribution and clearance of a material is critical for the medical treatment and long-term toxicity. The NPs solution (2 mg/mL, 0.2 mL) was injected intraperitoneally into C57 male mice (64 mice, 8 per group). Final dose in the mice was 20 mg/kg. The mice were scarified in 1, 7, 30, and 90 days, and main organs from 3 mice per group were tested by inductively coupled plasma mass spectrometry (ICP-MS). And then, the biodistributions of Bi from the PVP-protected $Bi_2Se_3$ NPs were obtained (Figure 2a). It is clear that liver, spleen, and kidney are the main target organs, and the bladder and testis are the next. Heart, lung, and brain showed relatively low Bi concentrations. Bi concentration was as high as 163.2 μg/g in liver, 101.9 μg/g in spleen, and 46.8 μg/g in kidney after 1 day. After 90 days, Bi concentration decreased to 9.26 μg/g in liver, 6.93 μg/g in spleen, and 0.78 μg/g in kidney. Biodistributions of Se in all organs are presented in Figure 2b. After 1 day treatment, Se also showed the highest uptake in liver, spleen, and kidney, with concentrations of 12.8, 10.9, and 4.25 μg/g, respectively. After 90 days, Se concentration decreased to 1.83 μg/g in liver, 0.23 μg/g in spleen, and 0.42 μg/g in kidney, respectively. The PVP layer prolonged the blood circulation time of the NPs.[53,54] It is not surprising that NPs larger than 10 nm can enter excretion system through the RES process. Phagocytosis by the Kupffer cells and spleen macrophages are significantly dependent on particle size. The larger particles can induce greater uptake by RES.[44,55] Further, it can be noticed that Bi concentration in all organs decreased with the increase of treatment days, which indicates the time-dependent clearance effects (Figure 2a). Se was detected mainly in liver, spleen, kidney, and bladder. Se concentration decreased in the organs until 30 days, but it increased in kidney, testis, and bladder after 90 days.

To investigate the clearance in detail, the time-dependent residual amount in mice from the injected $Bi_2Se_3$ NPs were investigated (Figure 2c,d). Residual amount of Bi and Se in mice were calculated by summing the Bi and Se amount in all organs from Figure 2a and 2b, respectively. The ICP-MS measurements showed that the initial total amount of Bi and Se was 190.76 and 124.68 μg, respectively. The residual ratios of Bi and Se were calculated by the residual amount / initial total amount. It was found that the residual amount of Bi in body after 1 day was 164.95 μg, and it was decreased to 11.66 μg after 90 days, which corresponds to 86.5 % of the initial amount after 1 day and 6.1 % after 90 days. Similar behavior was observed for Se. The residual amount was 13.43 μg after 1 day, and it decreased to 1.58 μg after 90 days, which indicated 10.8 % after 1 day and 1.3 % after 90 days. It is proposed that small NPs below 5.5 nm can be cleared by kidney, while large NPs above 50 nm may be partially cleared by liver as the time increased.[42] Thus, clearance by liver is a significant contribution to the decrease of the $Bi_2Se_3$ NPs in mice. Meanwhile, it is worth noting that high concentration of Bi and Se can also be found in kidney, indicating that kidney may participate in the partial metabolism of the NPs. It is similar to the previous work in which 50 nm Au particles showed unexpected high distribution in kidney.[44] Indeed, it was demonstrated that 10-30 nm graphene nanosheets could be cleared by both liver and kidney.[56,57] However, it is not clear whether metabolism of other organs is related to the breakdown of NPs *in vivo*. The NPs indeed can be degraded and excreted with long time blood circulation and physiological environment.[47]

It is notable that from the results in Figure 2, the Bi concentration was about 10 times higher than that of Se concentration. The as-synthesized $Bi_2Se_3$ NPs were stoichiometric, with a Bi/Se ratio of ~1.53 that was obtained from ICP-MS and energy dispersive spectrometer (EDS). The $Bi_2Se_3$ NPs were found to be oxidized easily in air environment. We found that the dry $Bi_2Se_3$ NPs powder was oxidized slowly to form $Bi_2O_{3-x}$, and pure Se was phase separated to form large aggregates. The same oxidation process was observed in the PBS buffer solution (pH=7.4). The $Bi_2Se_3$ powder suspended in the PBS buffer solution showed a slow oxidation process when they were exposed to air. We could not detect the trace of oxidation within a few days. However, severe oxidation was observed in one month. As seen in Figure 3a, the red amorphous Se was dissolved in the buffer solution and then precipitated on the glassware (indicated with an arrow). TEM image (Figure 3b) presents the shape deformation of the NPs into irregular amorphous-like particles which is due to the existence of amorphous Se (indicated with an arrow). The XRD peaks (Figure 3c) corresponds to a mixture of $Bi_2O_{2.33}$ and residual $Bi_2Se_3$, which confirms the replacement of Se by oxygen. Therefore, the as-synthesized NPs were preserved in Ar environment to prevent possible oxidation. The NPs before injection was identical in chemical composition and crystal structure to the as-synthesized ones.

The surprising mismatch of the atomic ratio before and after injection indicates fast oxidation of the NPs in mouse body. When the aqueous solution of the NPs was mixed with the blood plasma *in vitro*, the weight ratio of Bi/Se rapidly increased to 9.8 after 0.5 h, and further increased to 11.7 after 24 h (Figure 4a). These results confirm the fast oxidation of the NPs, which dissolves Se in mice blood. We further confirmed it by *in vivo* experiment. After *in vivo* injection, the Bi concentration sharply increased after 1 h and then decreased rapidly.

In contrast, the Se concentration reached a relatively stable level. The Bi/Se atomic ratio reached 5:1 after 1 h, but it decreased rapidly to 0.2 after 72 h (Figure 4b). Figure 4c schematically depicts the oxidation process in blood. The oxidized NPs or the pristine NPs were absorbed by RES, but Se atoms dissolved in blood remaining in the blood plasma. Probably existing in the complex forms with the biomolecules in the blood, Se could circulate for a long time in blood. Further, the Se dissolved in blood could be metabolized by renal clearance and thus accumulates in bladder.[42,58,59] Similar phenomenon has been reported in previous work where the PEG-coated gold NPs presented accumulation in organ after 90 days treatment.[45]

**2.3 Toxicity of the $Bi_2Se_3$ nanoplates**

We next investigated the *in vivo* toxicity of $Bi_2Se_3$ NPs at the dose of 20 mg/kg by evaluating body weight, immune response, hematology, and biochemistry at the time points of 1, 7, 30, and 90 days. During the 90 days period, the treatment with $Bi_2Se_3$ NPs did not cause obvious adverse effects on growth, and no meaningful statistical differences were observed in body weight and thymus index between the $Bi_2Se_3$ NPs-treated mice and control mice (Figure S2a and S2b). Spleen index in the NPs-treated mice increased steadily by 30 days, but it decreased after 90 days (Figure S2c). We selected standard hematology markers for analysis, including white blood cell (WBC), red blood cell (RBC), hematocrit (HCT), mean corpuscular volume (MCV), hemoglobin (HGB), platelet (PLT), mean corpuscular hemoglobin (MCH), and mean corpuscular hemoglobin concentration (MCHC). Hematology results for the $Bi_2Se_3$ NPs are presented in Figure 5a. For the NPs-treated mice, two typical indicators (WBC, RBC) have not shown meaningful differences from those of the untreated mice. PLT increased after 30 days treatment, but recovered after 90 days treatment. Other parameters (MCV, MCH, MCHC, HGB, HCT) had no significant difference. These results present limited biological damage to the mice. When the NPs were injected into the mice, they would interact with all serum protein, such as albumin, fibrinogen, γ-globulin, histone, and insulin. The blood chemistry of mice has not been affected directly. The PVP protected layer plays as a blood compatible surface coating layer, which is consistent with the previous works.[36,60] Compared with the Au-based nanomaterials at a similar dose range (10~20 mg/kg), the PVP-protected $Bi_2Se_3$ NPs caused less toxic responses. The PEG-coated Au NPs resulted in acute increase of RBC and WBC by intraperitoneal and oral injections because RBC and WBC are sensitive to infection and inflammation caused by the Au NPs.[43,45,59,61-63] These results show that the PVP-protected $Bi_2Se_3$ NPs do not cause significant infection and inflammation for mice.

Furthermore, we performed the standard biochemistry examination with the mice treated by the $Bi_2Se_3$ NPs at the different time points of 1, 7, 30, and 90 days (Figure 5b). The biochemical parameters including (a) alanine transaminase (ALT), (b) aspartate transaminase (AST), (c) total protein (TP), (d) albumin (ALB), (e) blood urea nitrogen (BUN), (f) creatinine (CREA), (g) globulin (GLOB), and (h) total bilirubin (TBIL) were investigated. We emphasize ALT, AST, and CREA because they are closely related to the functions of liver and kidney of mice. Meaningful changes were not observed in all parameters after 1 day. The NPs presented appreciable liver accumulation after 1 day, but did not cause acute injury at this time point. After 7 days treatment, ALT significantly increased in the NPs treated mice, while AST and CREA did not have noticeable difference in the treated mice. After 90 days, ALT has changed back to a normal level which indicates that the $Bi_2Se_3$ NPs caused slight damage for liver after 7 days, but the damage recovered after 30 and 90 days. It is consistent with the biodistribution and clearance of the NPs. It is widely reported that the liver accumulation of Au and metal oxide nanomaterials can induce continuous increase of ALT and AST, and thus significant damage in liver. Even at the dose level of 5 mg/kg through intravenous injection, Au nanoparticles and nanoclusters stabilized by PEG and BSA caused acute damage even after 90 days.[59,61] Similarly, the PEG- and Oleic acid-coated $Fe_3O_4$ nanocrystals with the size of 5-30 nm led to steady increase of ALT, AST, and BUN at the dose of 5-7.5 mg/kg.[64] However, the $Bi_2Se_3$ NPs showed very low liver toxicity even at 20 mg/kg dose.

Finally, we checked the pathological changes of organs by using immunohistochemistry at different time points. We collected liver, spleen, and kidney, and sliced them for Haematoxylin and Eosin (H&E) staining (Figure 6). No significant organ damages were observed in spleen and kidney during the whole treatment period. Slight pathological change was shown in liver after 7 days treatment, but it was recovered to the normal state after 90 days treatment. No noticeable toxic effects were detected in heart, lung, testis, bladder, and brain (Figure S3). Another feature of pathology is to qualitatively investigate the clearance of the $Bi_2Se_3$ NPs.[57,64] When the NPs are accumulated in an organ, they may aggregate and permit direct observations through optical microscope. A large number of dark spots (less than 1 μm) were found in spleen after 1 and 7 days treatment, but they were cleared after 90 days. Similar results were reported for carbon nanotube, graphene, and $Fe_3O_4$ NPs treated mice. Black spots were observed in liver, but they eventually disappeared after about 90 days.

Currently, nanomedicine is searching for materials with low toxicity and highly efficient clearance. It is widely accepted that small size nanoparticles can be cleared by kidney. Actual situation, however, is far more complex than the expectation. A case in point is about clearance of Au nanoparticles. The PEG-protected Au nanoparticles (3 nm) could not be cleared, but glutathione (GSH)-protected Au nanoparticles with the same size showed efficient renal clearance.[59,65] Carbon materials could be gradually cleared by kidney and liver when the size was up to 10~30 nm.[57] Therefore, it is shown that the clearance of nanomaterials is not only related to size and surface chemistry, but also related to the shape and stability. Exploring clearance of large size materials provides another available route for design of metabolizable nanomaterials. Present work clearly shows that the PVP-protected $Bi_2Se_3$ NPs can be metabolized by liver, hence they have large room for medical applications such as contrast agent and cancer therapy.

**2.4 Cancer radiation therapy of the Bi$_2$Se$_3$ nanoplates**

The cell responses (or biocompatibility) of the PVP-protected Bi$_2$Se$_3$ NPs were evaluated using the Hela cells. The Hela cell was treated by the Bi$_2$Se$_3$ NPs with different concentrations of 1.56 ~ 200 μg/mL. As shown in Figure 7a, the viabilities of the cells changed very little after 24 h and 48 h as the concentration of the NPs was increased, which indicates a low cytotoxicity even at a high dosage of 200 μg/mL. The *in vitro* radiation enhancement of the Bi$_2$Se$_3$ NPs was measured by the MTT assay using the Hela cells. As shown in Figure 7b, obvious enhancement in radiation with increasing dosage was observed for the cell cultures treated with the Bi$_2$Se$_3$ NPs. The radiation enhancement is regarded to be caused by increased DNA damage induced by the photoelectric effect and Compton scattering of the heavy metal (Bi).[66,67] This hypothesis was confirmed by the single-cell gel electrophoresis study. Without radiation, negligible DNA damage was observed for the cell treated with the NPs (Figure 7c). In contrast, after receiving 3 Gray (Gy) radiation dose, the DNA damage in the NPs-treated culture was much more significant than in the untreated culture (Figure 7c). The remarkable DNA damage was also confirmed by the *in vitro* imaging *via* the fluorescent DNA stain (Figure 7d). Long tails observed in the stain indicate significant DNA damage. There was no obvious DNA damage without radiation in both cells with/without NPs treatment, while long tails of stain were easily observed in the cells of the NPs-treated culture after 3 Gy radiations. It is clear that Bi$_2$Se$_3$ NPs is biocompatible and it can enhance DNA radiation damage. Radiation enhancement of the Bi$_2$Se$_3$ NPs is comparable with Au nanoparticles.[59]

Inspired by these results, we next investigated the *in vivo* biodistribution of the PVP-protected Bi$_2$Se$_3$ NPs using U14 tumor-bearing male nude mice (24 mice, 3 per group) after intraperitoneal injection (2 mg/mL, 0.2 mL) at the dose of 20 mg/kg. We used the inductively coupled plasma mass spectrometry (ICP-MS) to measure the concentrations of Bi and Se in blood at different time points of 0.5, 1, 2, 6, 12, 24, 48, and 72 h (Figure 8a). The concentrations of both Bi and Se sharply decreased in the first phase, and then gradually decreased in the second phase. In the first phase, the blood concentrations decayed to their half in 1.9 h for Bi and in 1.5 h for Se. The blood circulation times were lower that of 10-40 nm PEG-coated Au particles,[43,68,69] but higher than those of quantum dots.[32,68,70,71] Time-dependent concentrations of Bi and Se in tumor tissues were obtained from the same mice (Figure 8b). The Bi concentration in the tumor gradually increased from 0.35 μg/g (1.75 % ID/g) in 0.5 h to 1.48 μg/g (7.4 % ID/g) in 72 h, and then approached a plateau value in ~ 24 h. The concentration of Se in the tumor tissue increased from 0.25 μg/g in 0.5 h to 0.42 μg/g in 72 h. It is definite that the uptake in the tumor of the NPs is accumulated in tumor tissue by the passive targeting effect, which is consistent with previous works.[72,73] Further, it is found that the Bi concentration was close to the reported values of the Au nanoparticles (6.25 % ID/g) and the Au nanorods (7 % ID/g) accumulated in tumor tissues in 72 h after injection with the same dose.[68,69,71]

Encouraged by the highly effective accumulation of Bi in tumor cells, *in vivo* cancer radiation therapy was carried out by using U14 tumor-bearing nude mice. The PVP-protected $Bi_2Se_3$ NPs were intraperitoneally injected into mice (20 mg/kg). After 24 h, the mice were locally irradiated under $^{137}Cs$ gamma radiation of 3600 Ci at the dose of 5 Gy. The volume of the tumor was monitored by student T-test (Figure 8c). Without the radiation treatment, the tumor volume at the beginning steadily increased to 3.2 times larger after 25 days, regardless of the existence of the NPs. With the radiation treatment, the growth of the tumor volume in the mice without NPs injection (radiation only) was retarded by 6 days, but the tumor eventually grew to 2.4 times after 25 days. The tumor volume of the NPs-treated mice with radiation was 1.47 times after 25 days, which is a significant reduction in growth rate. It is notable that the tumor volume did not increase after 16 days until 25 days. Similarly, the tumor weight of the mice treated with NPs plus radiation exhibited a net increase of 0.1 g from the initial weight (0.21 g), which was less than the control mice and the mice with radiation only (Figure 8d). These *in vivo* results present that the PVP-protected $Bi_2Se_3$ NPs have a strong radiation enhancement effect and may be used as a radiosensitizer. Cancer radiation therapy of heavy metal elements mainly depends on the concentration in tumor and the photoelectric absorption. The photoelectric absorption coefficient of Bi is 0.12 $cm^2/g$ at radiation energy of 600 keV which is higher than that of Au (0.08 $cm^2/g$) at the same radiation energy. The effective accumulation in tumor tissue and the unique x-ray absorption induce obvious inhabitation of tumor growth after radiation.[74]

**2.5 X-ray CT imaging of the $Bi_2Se_3$ nanoplates.**

We next performed the *in vivo* X-ray CT imaging of the $Bi_2Se_3$ NPs in the mice bearing U14 tumor. The $Bi_2Se_3$ NPs (100 mg/ml, 0.2 ml) were intraperitoneally injected before anaesthesia. Three and two dimensions CT imaging were used to evaluate the tumor uptake and clearance (Figure 9). Three dimension imaging reveals the tumor outlines and boundaries of the tumor sites (indicated by the red circle) at 24 h after the injection of the $Bi_2Se_3$ NPs (Figure 9a). The $Bi_2Se_3$ NPs were also observed in the bladder (Figure 9 b), which indicates that the NPs were partially cleared by kidney. Figure 9c-e show time-dependent two dimension CT imaging of the $Bi_2Se_3$ NPs. The tumor boundary was not clear at 3 h, but it was discriminated gradually as the time increased to 24 h, which is attributed to the passive accumulation of the NPs in tumor. Figure 9f-h show the bladder. The bladder was seen clearly at 3 h and started to lose its intensity after 24 h, which indicates the time-dependent renal clearance effect. It was reported that small particles (<5.5 nm) can be cleared by kidney.[42,58,59] Our work, however, suggest the $Bi_2Se_3$ NPs can also be metabolized partially by renal clearance. The X-ray CT imaging is related to atom number and relative ratio of organ/muscle. Au nanoparticles were reported to be used as X-ray contrast agent at the concentration of 100 mg/g.[75,76] In our work, X-ray CT imaging with the $Bi_2Se_3$ NPs could exhibit the clear boundary between tumor and normal tissue at the similar concentration with the Au nanoparticles.

## 3. Conclusion

This study explored the distribution of the $Bi_2Se_3$ nanoplates (NPs) (53.8 in width, 6 nm in thickness) in the U14 tumor-bearing male mice. The NPs stabilized with PVP exhibited a long blood circulation half-life (>50 h). We investigated the *in vivo* toxicity, biodistribution, and clearance of the PVP-protected $Bi_2Se_3$ NPs at the dose of 20 mg/kg. It was found that liver and spleen were main target organs. Large amount of Bi were found in liver and spleen after 1 day treatment (~163.2 μg/g in liver, ~101.9 μg/g in spleen), but more than 93 % of Bi were cleared after 90 days treatment. The concentrations of Bi and Se in the organs significantly decreased after 25 days, but the accumulation of Bi in the tumor cells increased through the enhanced permeability and retention (EPR) of the cancerous tumors. The Bi elements accumulated in the tumor cells were employed as a radiosensitizer to inhibit the growth of the cancer cells and enhance the contrast in the computerized tomography (CT) imaging. Once treated with the NPs and radiation, the growth of the cancer cells was significantly delayed and the cancer cells did not further grow after 16 days. This work suggests a new way of radiation cancer therapy employing the $Bi_2Se_3$ nanoplates. The effect of diverse surface stabilizers on the circulation time and biodistribution of Bi in the tumor cells is left for the future study.

## 4. Experimental section

*Materials and synthesis*: 1.0 g of poly(vinylpyrrolidone)(PVP, $M_w$=55 000, Sigma-Aldrich) was dissolved in 50 ml of ethylene glycol (EG, ≥ 99 %, J. T. Baker), and the solution was poured in 250 ml round-bottom flask. A solution of sodium selenite (99 %, Aldrich, 0.242 g in 40 ml of EG) and a solution of bismuth nitrate pentahydrate (99.99+ %, Aldrich, 0.452 g in 15 ml of EG) were added to the PVP solution under magnetic stirring at room temperature. The flask was sealed and heated to 180 °C in nitrogen environment. As the reaction temperature increased, transparent reaction mixture became milky white and turned yellow-white. Hydrazine hydrate solution (2 ml in 20 ml of EG) was rapidly injected to the mixture by a syringe. The reaction mixture turned dark immediately, which indicates the formation of the PVP-protected $Bi_2Se_3$ NPs. The reaction was allowed to proceed for 30 min for a complete reaction and cooled down to room temperature. The final products were precipitated by centrifuging (12,000 rpm, 10 min) and washed three times with a mixture of acetone (300 mL) and D.I. water (60 mL). And, the precipitate was washed with pure DI water twice and dried in vacuum.

*Material characterization*: Transmission electron microscope (TEM) analysis was conducted at 200 kV with JEOL models (JEM-2100F and JEM-2011HC). X-ray diffraction (XRD) measurement was performed with a Rigaku D/MAX II x-ray diffractometer at Cu $K_\alpha$ radiation ($\lambda = 0.1542$ nm). The scanning range extended from 10 to 80° with a speed of about 3°/min. The zeta-potential was measured by electrophoretic measurements (ELS-Z2). Size distribution of the $Bi_2Se_3$ NPs was determined with the NanoZS Zetasizer particle analyzer (Malvern). Sample solutions were prepared by diluting $Bi_2Se_3$ NPs into 10 mM PBS solution (pH 7.0) and data were acquired in the phase analysis light scattering mode at 25 °C. Stability of the $Bi_2Se_3$ NPs was also evaluated using light scattering. $Bi_2Se_3$ NPs were diluted five times in PBS, and size distribution was measured after 3 month in a 5 ml glass cuvette to check the possible change in size.

*Biodistribution*: The organs and the solutions from the $Bi_2Se_3$ NPs-treated mice were digested by using a microwave system CEM Mars 5 (CEM, Kamp Lintfort, Germany). The Bi and Se content were measured with an ICP-MS, type Agilent 7500 CE (Agilent Technologies, Waldbronn, Germany).

*Hematology, biochemistry, and pathology*: Using a standard saphenous vein blood collection technique, blood was drawn for hematology analysis (potassium EDTA collection tube). The analysis of standard hematological and biochemical examination was carried out.

For blood analysis, 1 mL of blood was collected from mice and separated by centrifugation into cellular and plasma fractions. Mice were sacrificed by isoflurane anesthetic and angio catheter exsanguinations, and major organs from those mice were harvested, fixed in 10% neutral buffered formalin, processed routinely into paraffin, stained with H&E and pathology were examined by a digital microscope.

*In vitro Cytotoxicity Test*: Hela cells were cultured at 37 °C in humidified atmosphere with 5% $CO_2$ and high-glucose Dulbecco's modified Eagle's medium (DMEM) which contained fetal calf serum (10%), L-glutamine (2.9 mg/ml), streptomycin (1 mg/ml), and penicillin (1000 units/ml). The cells (in culture medium) were dispensed in 96-well plates (90 μl containing 6000 cells per well). Different concentrations of the $Bi_2Se_3$ NPs (10 μl) were then added to each well. The effect of the concentration of $Bi_2Se_3$ NPs was assessed using MTT Cell Proliferation and Cytotoxicity Assay Kit. After 24 or 48 hours of treatment, 10 μl of MTT reagent was added and incubate for 4 hours, then the medium were replaced with 150 μl DMSO. The optical absorption in 490 nm signal was recorded with a single tube luminometer (TD 20/20, Turner Biosystems Inc., Sunnyvale, CA, USA).

*In Vivo Toxicity*: 64 male C57 mice at 11 weeks of age were obtained from the Institute of Radiation Medicine (IRM) laboratories and were housed by 2 mice per cage in a 12 h/12 h light/dark cycle, and were provided with food and water *ad libitum*. Mice were randomly divided into eight groups (eight mice in each group): 1 day control, 1 day treated, 7 days control, 7 days treated, 30 days control, 30 days treated, and 90 days control, 90 days treated mice, respectively. A $Bi_2Se_3$ NPs solution (200 μL, 2 mg/mL) was used for the animal experiment using intraperitoneal injection. The concentration was 20 mg/kg in each mouse. The animals were weighed and assessed every day to check the behavioral changes. After 1, 7, 30, and 90 days treatment, all mice were sacrificed, and blood and organs were collected for biochemistry and pathological studies. Mice were sacrificed using isoflurane anesthetic and angiocatheter exsanguination with PBS. One mouse from each group was fixed with 10% buffered formalin following phosphate-buffered saline exsanguination. During necropsy, liver, kidneys, spleen, heart, lung, testis, brain, bladder, and thymus were collected and weighed. To explicitly examine the grade of changes caused by malities, spleen and thymus indexes ($S_x$) were used:

$$S_x = \frac{\text{Weight of experimental organ } (mg)}{\text{Weight of experimental animal } (g)}$$

*In vitro DNA Break*: A modified version of the alkaline COMET-assay protocol was performed to assay the DNA break. In a typical assay, frosted microscope slides were covered with 200 μL of 0.1% agarose in PBS. After the solidification of agarose, $2\times10^5$ cells suspended in 10 μL of PBS and 75 μL of 0.5% low-melting-point agarose were added to each slide. After solidification, the slides were placed in cold fresh lyses buffer [2.5 M NaCl, 100 mM disodium ethylenediaminetetraacetate (EDTA), 10 mM Tris-HCl, and 1% Triton X-100] for 1 h and subsequently in a horizontal gel electrophoresis unit (20 × 25 cm) filled with chilled electrophoresis buffer (300 mM NaOH and 1 mM $Na_2EDTA$) for 30 min. Electrophoresis was then conducted at 14 V for 1 h. The slides were drained, neutralized, and dried with ethanol after the electrophoresis. The comets were stained with ethidium bromide. The DNA damage was analyzed using Comet Assay Software Project (CASP) software that measures the tail moment.

*In vitro Radiation Therapy*: Hela cells were incubated in 96-well plates (6000 per well) overnight and then exposed to the $Bi_2Se_3$ NPs (100 μl, 200 μg/mL) for another 24 h. Then, the cells were irradiated under gamma-rays from $^{137}Cs$ (photon energy 662 keV) with an activity of 3600 Ci at the doses of 1, 2, 4, 6, and 8 Gy. After 24 hour of irradiation, cells viability was measured by MTT assay.

*In Vivo Radiation Therapy*: All animals were purchased, maintained, and handled using protocols approved by the Institute of Radiation Medicine (IRM) at the Chinese Academy of Medical Sciences (CAMS). The U14 tumor models were generated by subcutaneous injection of $2\times10^6$ cells in 50 μL PBS into the right shoulder of female BALB/c mice. The mice were treated with the PVP-protected $Bi_2Se_3$ NPs when the tumor volume reached 100 to 130 mm$^3$ (7 days after tumor inoculation). For control, 200 μL saline was injected into mice via intraperitoneal injection. For the treatment, 2 mg/mL of the $Bi_2Se_3$ NPs was used for the animal experiment, and the concentration was raised up to 20 mg/kg in the mice. Subsequently, the mice were radiated by 5 Gy gamma rays, and $^{137}$Cs with activity of 3600 Ci and photon energy of 662 KeV were used. 32 mice were assigned to the following groups (eight mice in each group): control, the PVP-protected $Bi_2Se_3$ NPs, radiation alone, the $Bi_2Se_3$ NPs plus radiation. The tumor size was measured every two or three days and calculated as: volume = (tumor length) × (tumor width)$^2$/2.

*Statistical analysis*: All data presented in this study are the average ± SD of at least three or more experimental results. The paired Student's t-test was used for the statistical analysis.

**Acknowledgements**

This work was supported by the National Natural Science Foundation of China (Grant No.81000668), Natural Science Foundation of Tianjin (Grant No. 13JCQNJC13500), the Development Foundation of IRM, CAMS (Grant No.SF1207, 1336) and Union New Star, CAMS (No.1256). U.J. acknowledges the National Research Foundation (NRF) grant funded by the Korean Government (MEST) through the World Class University Program (R32-20031).

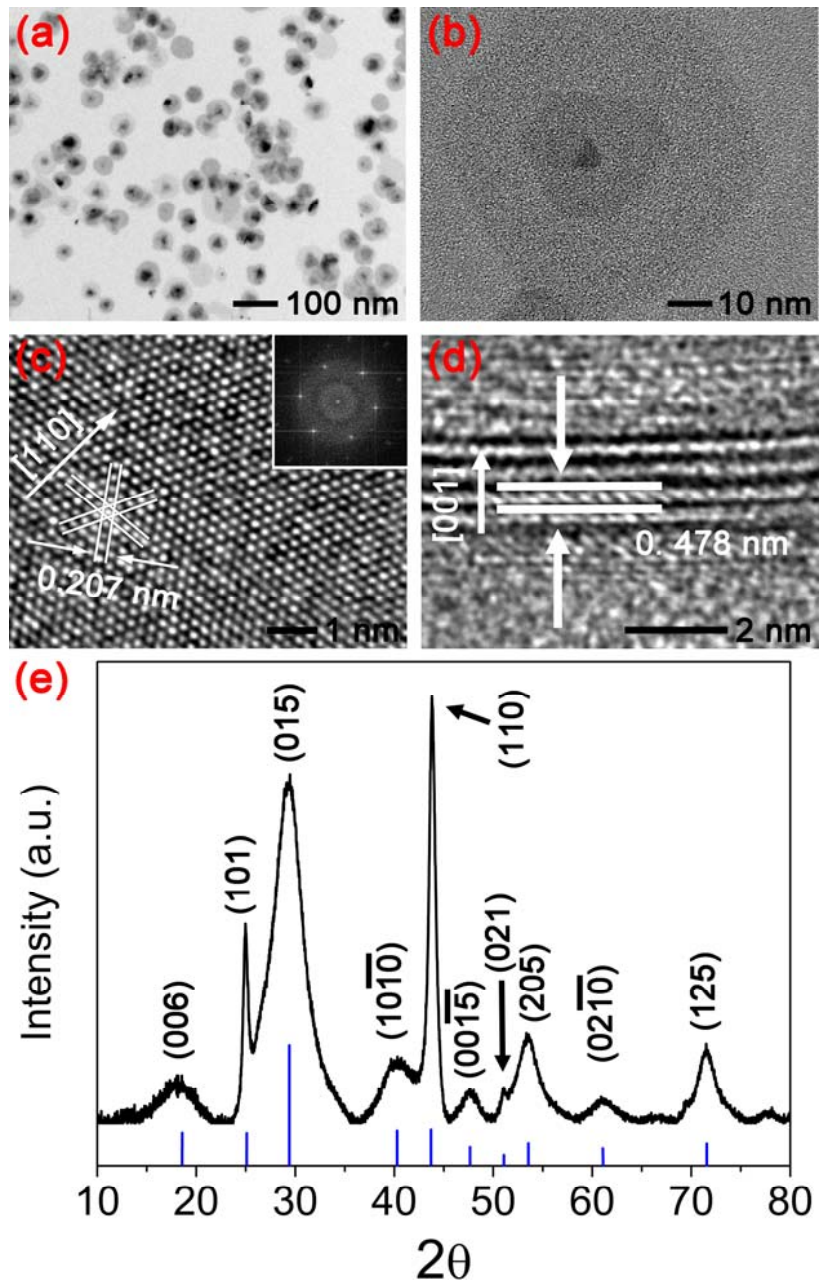

**Figure 1.** (a) Representative TEM image of the PVP-protected $Bi_2Se_3$ nanoplates (NPs), and (b) a magnified image showing a terraced 2-D structure of the NPs. (c,d) High-resolution TEM image of the as-prepared PVP-protected $Bi_2Se_3$ NPs in (c) horizontal plane and (d) in thickness direction. (e) XRD image of the as-prepared PVP-protected $Bi_2Se_3$ NPs. The blue lines are the peaks of bulk $Bi_2Se_3$.

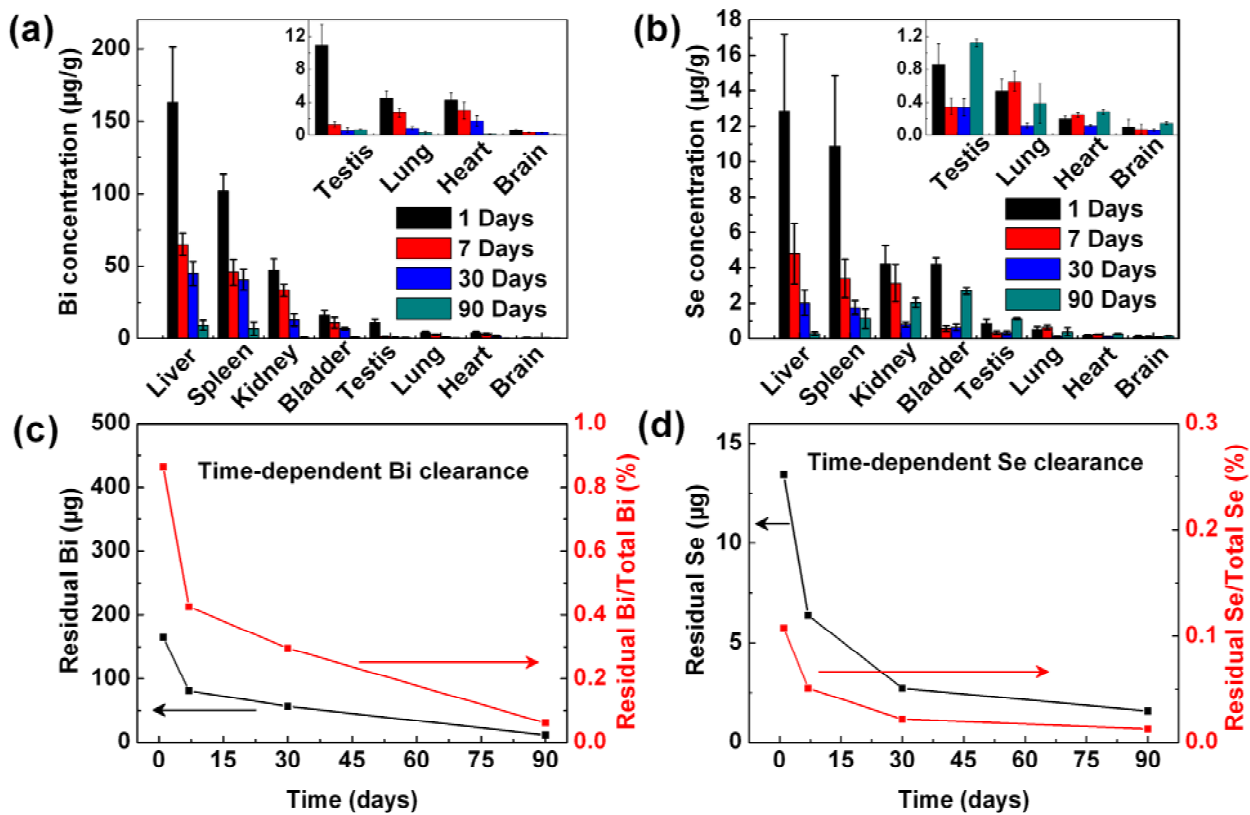

**Figure 2.** *In vivo* biodistribution and clearance of the PVP-protected $Bi_2Se_3$ NPs. (a) Bi concentration and (b) Se concentration at the different time points of 1, 7, 30, and 90 days after intraperitoneal injection (20 mg/kg). (c,d) Residual amount of Bi (c) and Se (d), and the corresponding atomic ratio to the initial atomic amount from the injected $Bi_2Se_3$ NPs.

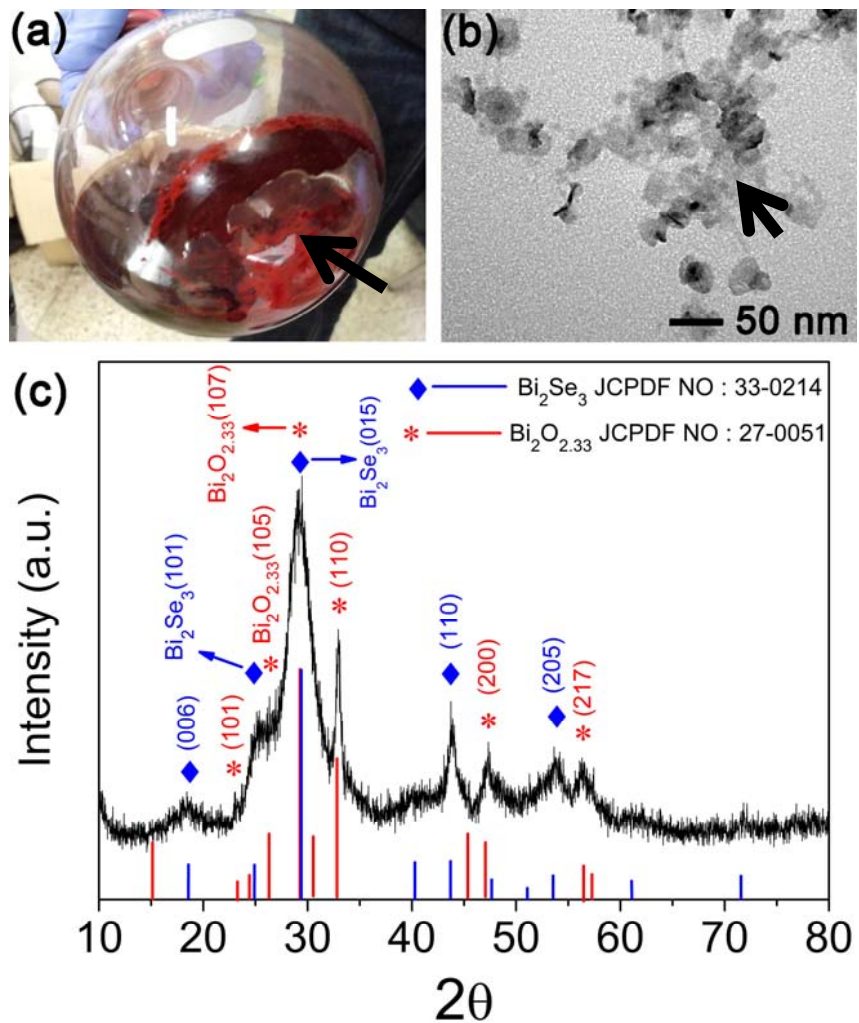

**Figure 3.** Oxidation of the $Bi_2Se_3$ NPs in a PBS buffer solution exposed to air for 30 days. (a) Digital image of the NPs suspension. The red colored material precipitated on the bottom surface of the flask is amorphous Se (indicated with an arrow), (b) TEM image showing deformation of the NPs by the oxidation (indicated by an arrow), and (c) XRD spectrum of the NPs collected from the PBS buffer solution by vacuum filtration. The red vertical lines and the blue vertical lines at the bottom present XRD peaks of pure $Bi_2Se_3$ and $Bi_2O_{2.33}$.

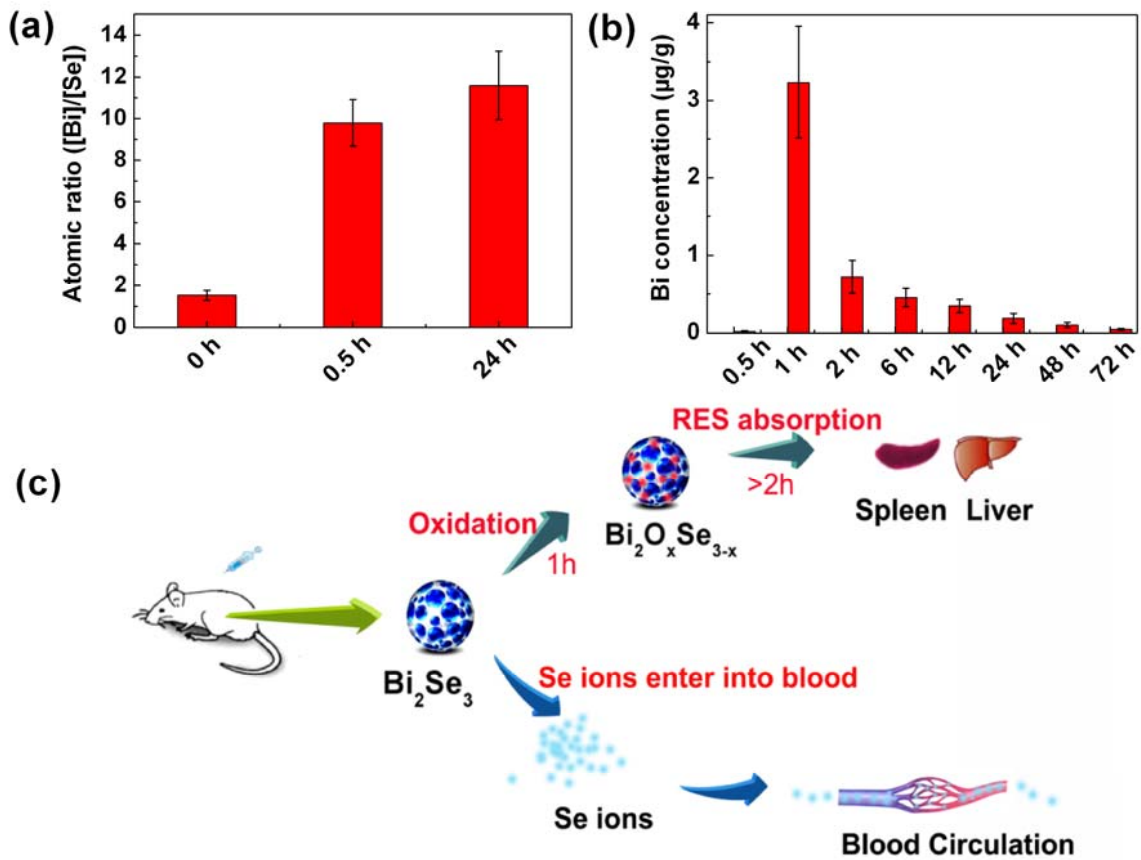

**Figure 4.** Oxidation process of Bi$_2$Se$_3$ NPs. (a) Time-dependent atomic ratio of Bi to Se *in vitro*. (b) Time-dependent concentration of Bi which was measured by ICP-MS *in vivo*. (c) Illustration to describe the dissolution and oxidation of the Bi$_2$Se$_3$ NPs after intraperitoneal injection.

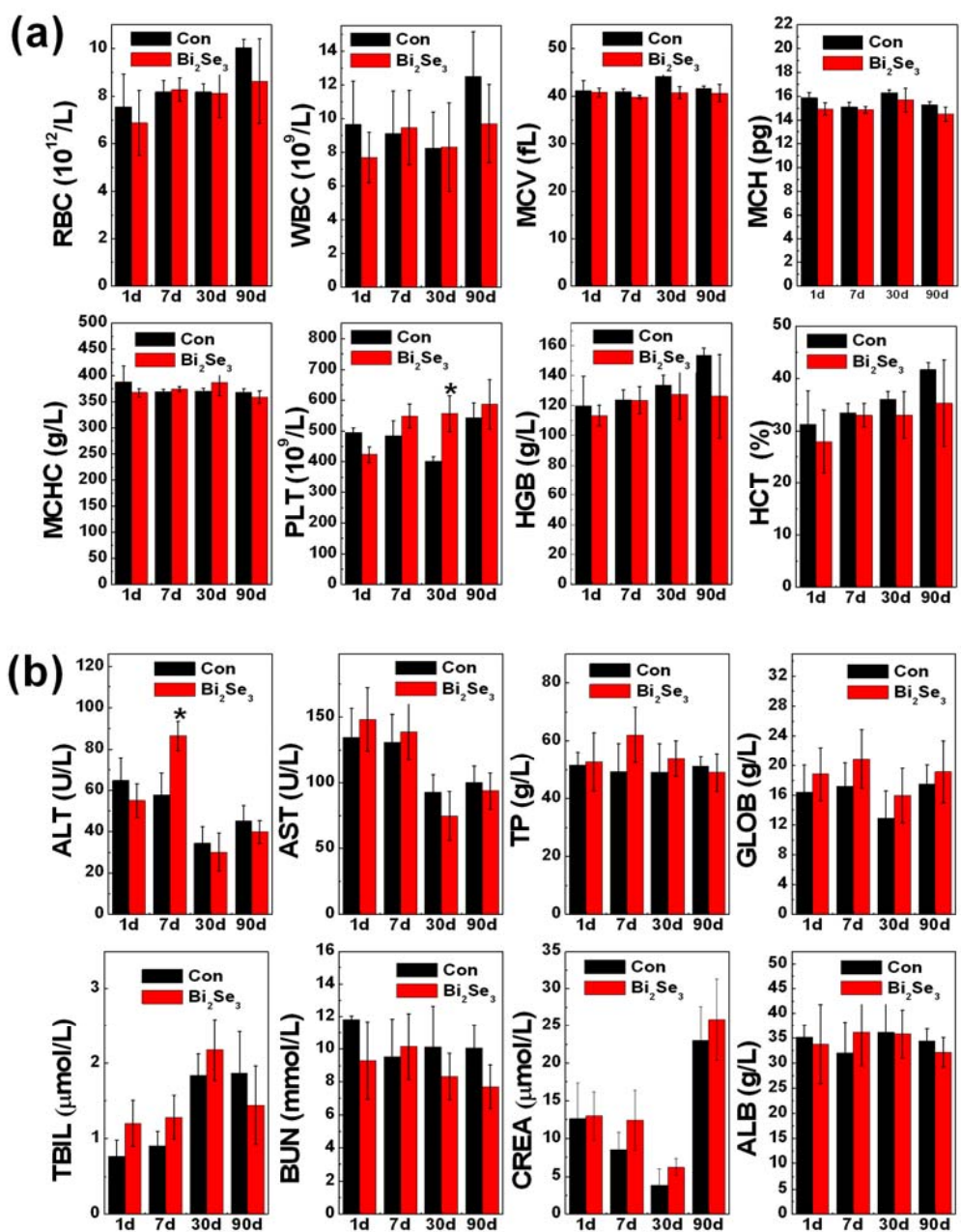

**Figure 5.** (a) Hematology data of the mice treated with the PVP-protected $Bi_2Se_3$ NPs. The data were collected at the different time points of 1, 7, 30, and 90 days after intraperitoneal injection (20 mg/kg). The terms are following: red blood cells (RBC), white blood cells (WBC), platelets (PLT), mean corpuscular hemoglobin (MCH), mean corpuscular hemoglobin concentration (MCHC), mean corpuscular volume (MCV), hemoglobin (HGB), and hematocrit (HCT). The terms 'Con.' and '$Bi_2Se_3$' in the figures indicate controlled mice without the $Bi_2Se_3$ NPs and the mice treated with the NPs, respectively. (b) Blood biochemistry analysis of the mice treated with the PVP-protected $Bi_2Se_3$ NPs at the different time points of 1, 7, 30, and 90 days. The results show mean and standard deviation of aminotransferase (ALT), aminotransferase (AST), total protein (TP), albumin (ALB), blood urea nitrogen (BUN), creatinine (CREA), globulin (GOLB), and total bilirubin (TBIL). Data were obtained by Student's t-test.

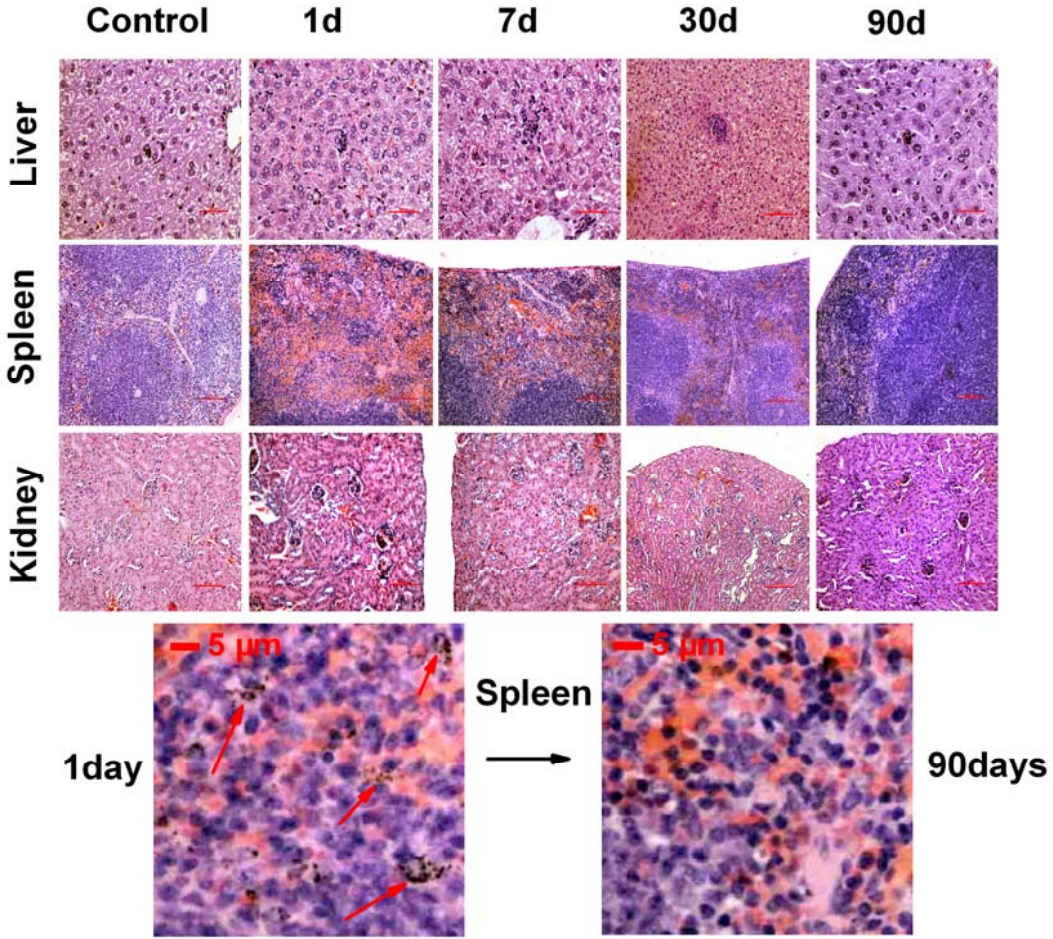

**Figure 6.** Pathological data from the liver, spleen, and kidney of the mice treated with the PVP-protected $Bi_2Se_3$ NPs. The data were collected at different time points of 1, 7, 30, and 90 days after injecting 20 mg/kg dose.

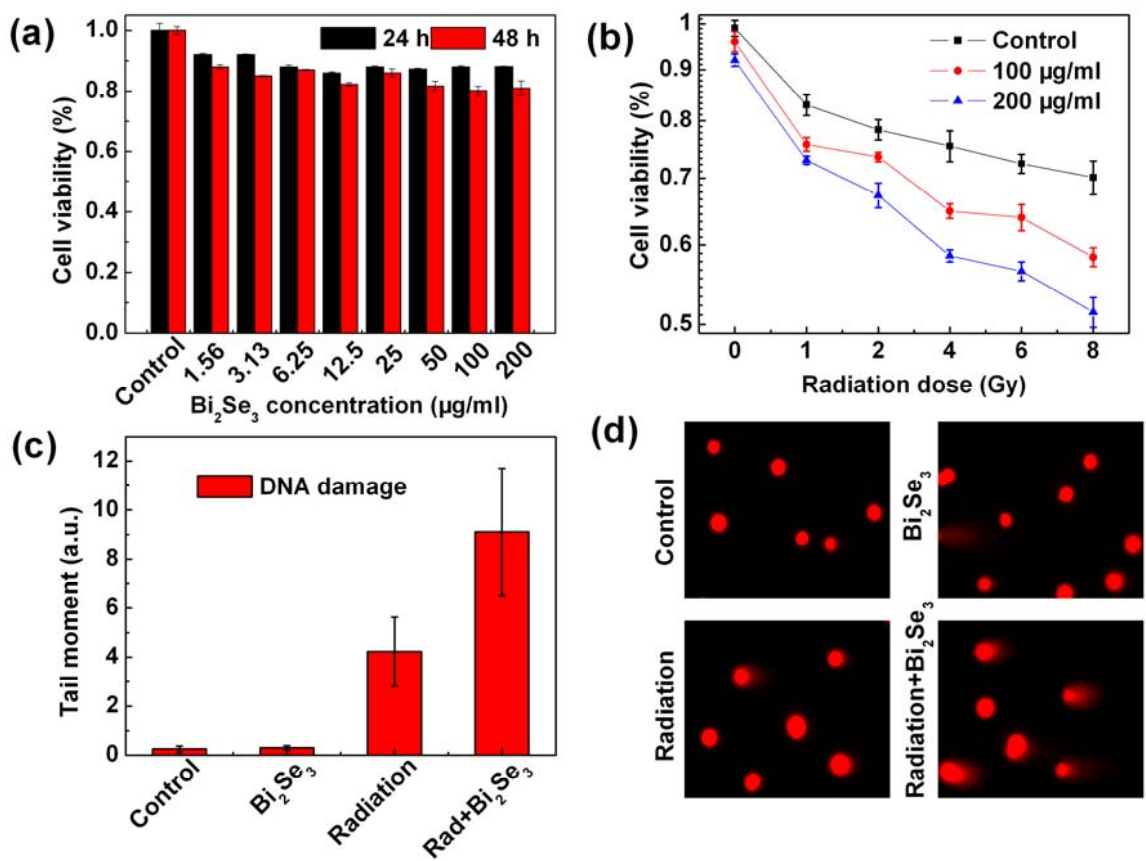

**Figure 7.** (a) Viability of Hela cells after incubation with the PVP-protected $Bi_2Se_3$ NPs for 24 and 48 h. (b) Viability of Hela cells treated with only radiation [control, black line], the $Bi_2Se_3$ NPs (100 μg/mL) + radiation [red line], and the $Bi_2Se_3$ NPs (200 μg/mL) + radiation [blue line]. (c) Tail moment of Hela cells treated with the $Bi_2Se_3$ NPs, radiation (3 Gy), and the $Bi_2Se_3$ NPs + radiation (3 Gy). (d) Representative cell images of fluorescent DNA stain of control, radiation group, the $Bi_2Se_3$ NPs, and the $Bi_2Se_3$ NPs + radiation (3 Gy).

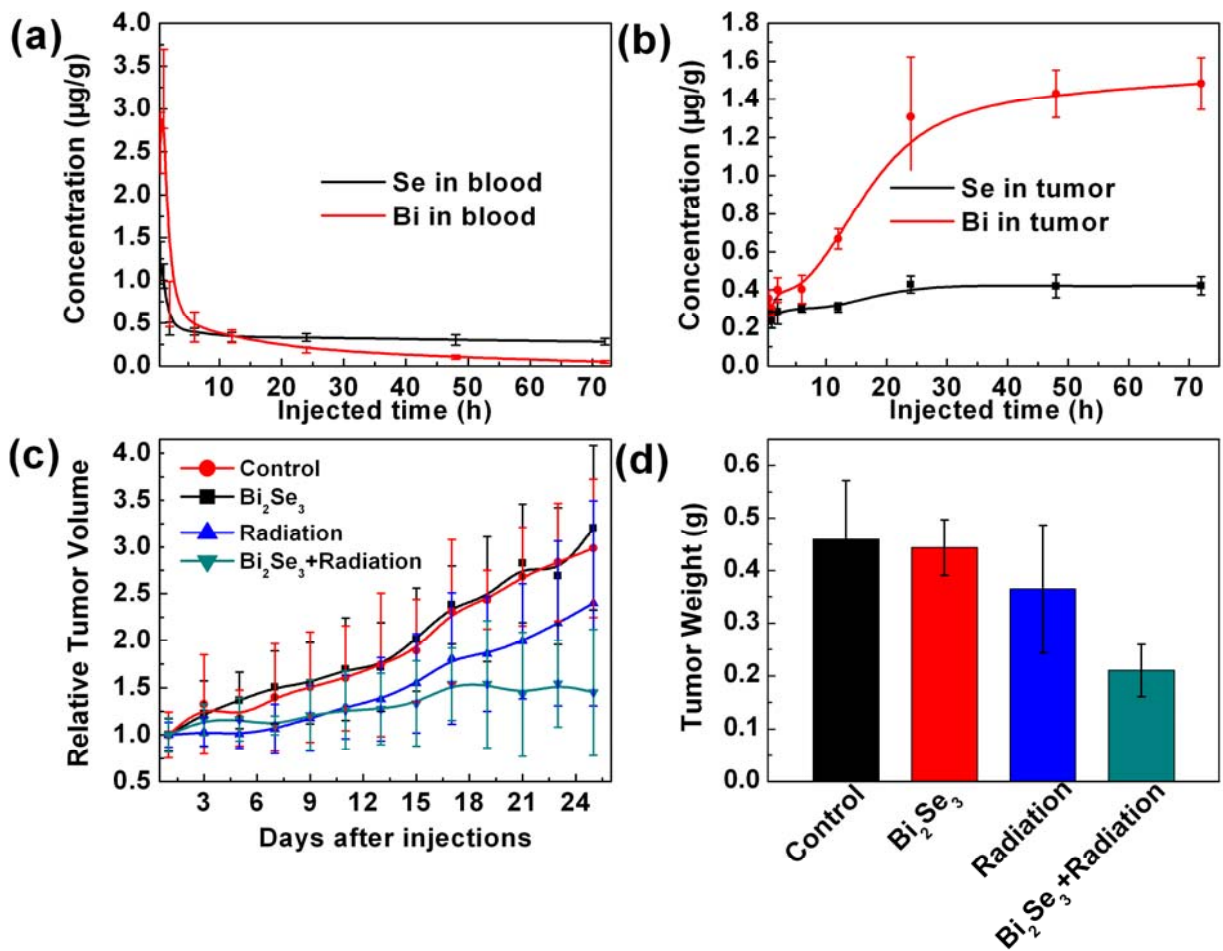

**Figure 8.** *In vivo* biodistribution and passive targeting of the PVP-protected $Bi_2Se_3$ NPs. Concentration of Bi and Se in blood (a) and in tumor (b) at the different time points of 0.5, 1, 2, 6, 12, 24, 72 h after intraperitoneal injection (20 mg/kg). Time-course studies of the tumor volume (c) and the tumor weight (d) of mice treated with the PVP-protected $Bi_2Se_3$ NPs at the concentration of 20 mg/kg. Data was obtained by Student's t-test.

 

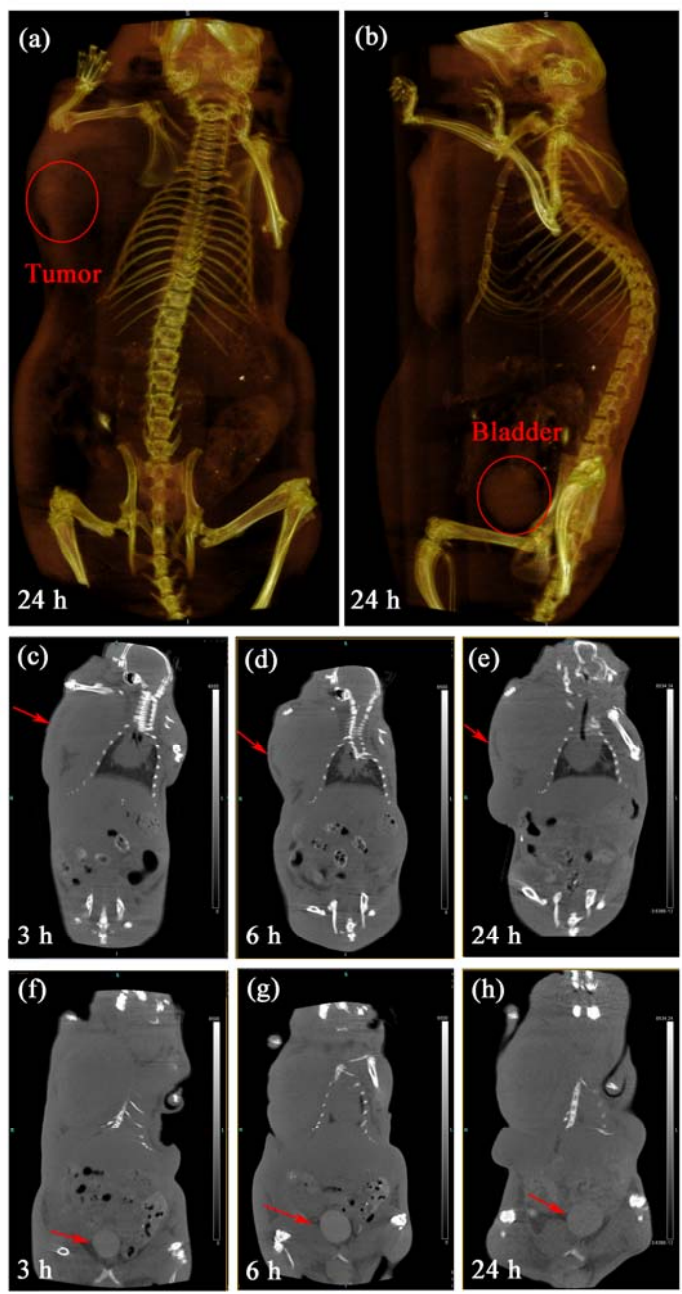

**Figure 9.** Three dimension *in vivo* X-ray CT imaging of tumor tissue (red circle) and bladder (red circle) from the PVP-protected $Bi_2Se_3$ NPs from front view (a) and from side view (b) at 24 h. (c-e) Two dimension images of the tumor tissue (red arrow) were taken at (c) 3 h, (d) 6 h. and (e) 24 h. (f-h) Two dimension images of bladder (red arrow) from PVP-protected $Bi_2Se_3$ NPs treated mice at (f) 3 h, (g) 6 h, and (h) 24 h.

# **Supporting Information**

## Metabolizable $Bi_2Se_3$ Nanoplates: Biodistribution, Toxicity, and Uses for Cancer Radiation Therapy and Imaging

*Xiao-Dong Zhang,[†] Jie Chen,[†] Yuho Min,[‡] Gyeong Bae Park,[‡] Xiu Shen,[†] Sha-Sha Song,[†] Yuan-Ming Sun,[†] Hao Wang,[†] Wei Long,[†] Jianping Xie,[§] Kai Gao,[∥] Lianfeng Zhang,[∥] Saijun Fan,[†] Feiyue Fan,[†] and Unyong Jeong*[‡]*

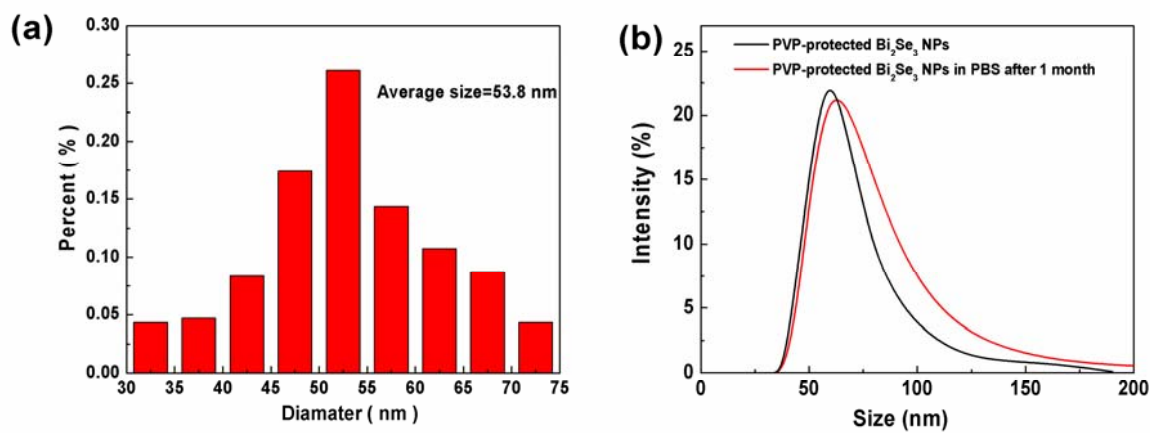

**Figure S1.** Size distribution of the as-prepared PVP-protected $Bi_2Se_3$ NPs obtained from TEM images (a), and from dynamic light scattering spectra in PBS buffer solution (b). Data in (b) contain the initial size and the value after 1 month.

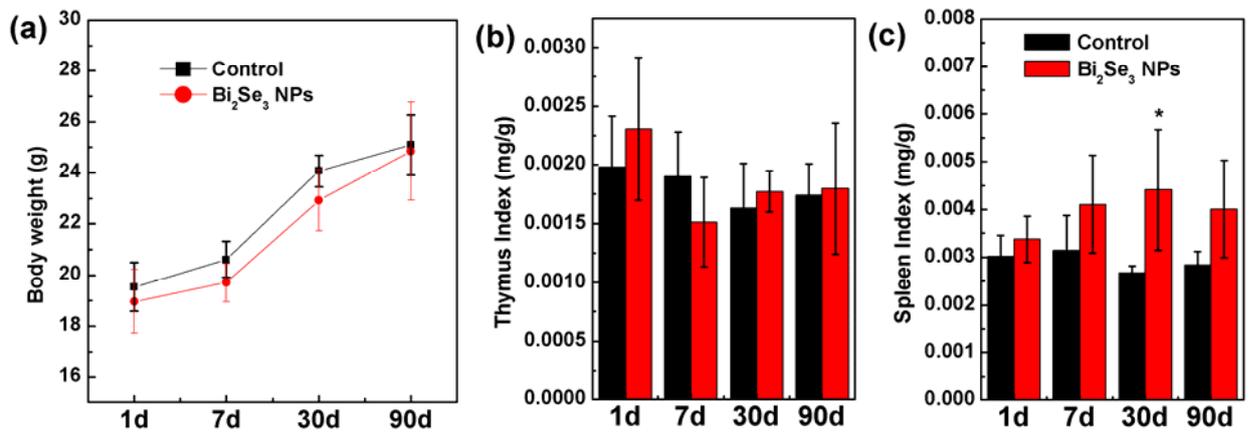

**Figure S2.** (a)The body weight, (b) thymus index, and (c) spleenindex of the mice treated with the PVP-protected $Bi_2Se_3$ NPs.

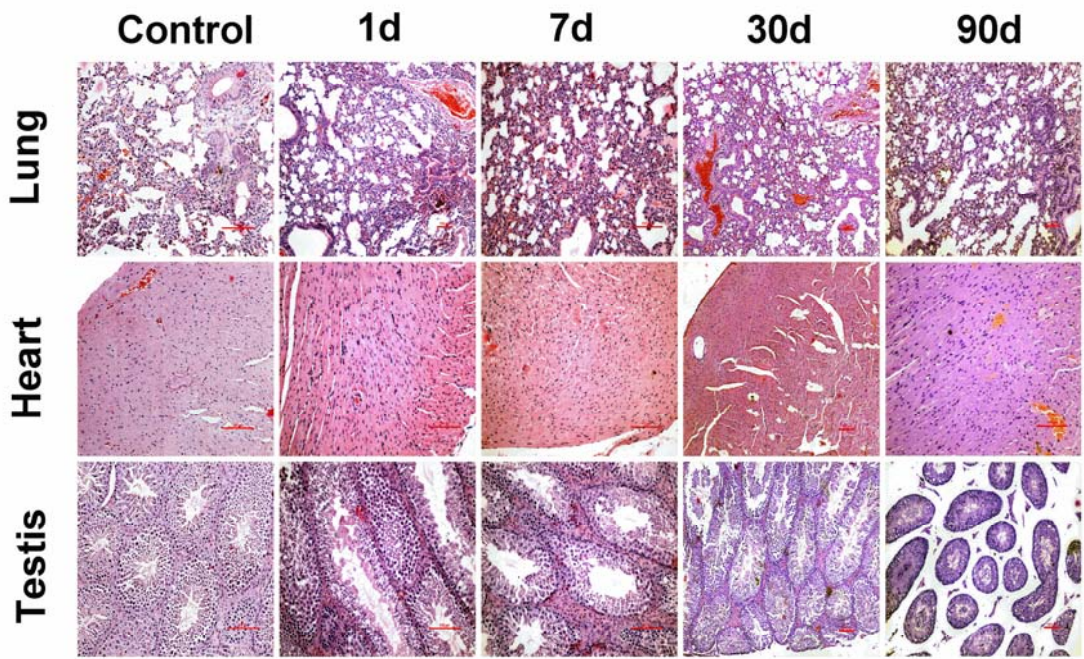

**Figure S3.** Pathologicalimage of the heart, lung,and testis of the control mice group (untreated) and of the mice treated with the PVP-protected $Bi_2Se_3$ NPs**.**